\documentclass[letterpaper]{IEEEtran}
\usepackage{dblfloatfix}
\usepackage{amssymb}
\usepackage{amsmath, amsthm}
\usepackage{amsfonts} 
\usepackage{fancyhdr}
\usepackage{cite}
\usepackage{graphicx, wrapfig, subcaption, setspace, booktabs}
\usepackage{caption}
\usepackage[protrusion=true, expansion=true]{microtype}
\usepackage[english]{babel}
\usepackage{url, lipsum}
\usepackage{mdwmath}
 \usepackage{color,soul}
 
 \usepackage{algorithm}
\usepackage{algpseudocode}
\usepackage[T2A, T1]{fontenc}
\usepackage[utf8]{inputenc} 
\usepackage{amsmath}
\usepackage{mathrsfs}

\ifCLASSINFOpdf

\else

\fi

\usepackage{amsmath}

\begin{document}
\title{Block Phase Tracking Reference Signal (PTRS) Allocation for DFT-s-OFDM }

\author{Mostafa~Ibrahim,~
        Sabit Ekin,~\IEEEmembership{Senior~Member,~IEEE},
        Yingming (Allan) Tsai,~\IEEEmembership{Senior~Member,~IEEE}, and~Ravikumar~Pragada,~\IEEEmembership{Senior~Member,~IEEE}

\thanks{        (\textit{*Corresponding author: Mostafa Ibrahim}.)}

\thanks{M. ~Ibrahim, Allan, and R.~Pragada are with InterDigital Communications. Inc. PA, USA, (E-mail: mostafa.ibrahim,  yingming.tsai, ravikumar.pragada\{@interdigital.com\}).}
\thanks{M. ~Ibrahim is with the Department of Engineering Technology, Texas A\&M University, Texas, USA (E-mail: mostafa.ibrahim@tamu.edu).}
\thanks{S.~Ekin  is with the Departments of Engineering Technology, and Electrical \& Computer Engineering, Texas A\&M University, Texas, USA (E-mail: sabitekin@tamu.edu).}
}
\maketitle
\begin{abstract}


This study introduces a Block Phase Tracking Reference Signal (PTRS) allocation approach for Discrete Fourier Transform-spread-Orthogonal Frequency Division Multiplexing (DFT-s-OFDM) systems to enhance phase noise tracking and compensation. 
Our proposed block allocation methodology leverages the concepts of multiresolution time-frequency tiling for more effective sampling, thereby mitigating aliasing effects and improving phase noise resilience.  A key contribution of our approach is a novel modulation and demodulation scheme, incorporating a dedicated DFT-s-OFDM symbol, a modulator branch for block PTRS generation, and a dedicated demodulator for accurate phase noise estimation and correction. 
Our findings demonstrate substantial improvements in DFT-s-OFDM waveform performance in comparison with 5G New Radio (NR), promising enhanced efficiency and reliability for future mobile communication systems.
\end{abstract}


\IEEEpeerreviewmaketitle

\section{Introduction}
\IEEEPARstart{S}{\lowercase{ingle}}-carrier schemes are anticipated to gain prominence in future mobile communications, notably in 6G and beyond. This shift is driven by the adoption of millimeter wave (mm-wave) and Terahertz technologies, enabling operations in higher frequency domains where power amplifiers exhibit reduced efficiency, and the peak-to-average power ratio (PAPR) of transmitted waveforms emerges as a critical concern \cite{designBeyond6g}. Moreover, the evolution towards high-speed line-of-sight (LOS) backhaul-like scenarios, characterized by less frequency selectivity, positions single-carrier schemes as advantageous due to their inherently low PAPR and aptness for low-frequency selectivity environments. 
A prominent single-carrier scheme, Single-carrier Frequency Division Multiple Access (SC-FDMA), is utilized in the 4G Long Term Evolution (LTE) and 5G-New Radio (NR) standards. SC-FDMA, also referred to as Discrete Fourier Transform-spread-OFDM (DFT-s-OFDM) \cite{demir2019waveform} is a baseline for many single-carrier-based waveform variations \cite{berardinelli2013zero,sahin2015improved,ibrahim2016zero,10123918,10356370}.

DFT-s-OFDM faces phase noise and asynchronicity challenges \cite{Raghunath} since the modulated data symbols are sequential pulses in the time domain rather than subcarriers in the frequency domain, as seen in Orthogonal Frequency Division Multiplexing  (OFDM). Therefore, phase noise affects SC-FDMA differently from OFDM, necessitating distinct tracking and compensation strategies. Particularly, DFT-s-OFDM is more prone to this issue in the future generations of mobile communications because the impact of phase noise grows higher as the carrier frequency increases \cite{lomayev2020method, lee2000oscillator}, where it is envisioned to operate. Moreover, higher-order modulation is severely limited by the phase noise effect \cite{khanzadi2013estimation}, which is throughput limiting.

Phase Locked Loops (PLLs) are commonly employed for local oscillators and frequency synthesizers \cite{gardner2005phaselock}. Phase noise is a type of hardware impairment that results from various factors, such as thermal noise added to the voltage-controlled oscillator (VCO) and non-integer division factors in the PLL feedback decimator. It follows a Brownian motion random process and is typically modeled as a wide-sense stationary process. Phase noise is generally described by its frequency domain power spectral density (PSD) in dBc/Hz, which is the ratio between the noise power in a bandwidth of 1 Hz and the power of the carrier at a specific center frequency, as described in \cite{hajimiri1998general}. Several studies have used multi-pole/zero models to describe the phase fluctuations \cite{wang2017low}.

The impact of phase noise on SC-FDMA differs from its impact on OFDM. In OFDM, phase noise causes a common phase shift and intercarrier interference (ICI).
The common phase shift affects all symbols, while ICI causes interference between symbols due to the convolution effect in the frequency domain. Phase noise and ICI compensation in OFDM  has been studied in \cite{choi2023phase,afshang2022phase,mathecken2017constrained}. 
The effect of channel selectivity added to phase noise results in inter-symbol interference (ISI). In the literature, there are studies on iterative \cite{syrjala2012iterative,linsalata2023joint} and non-iterative \cite{syrjala2019pilot,rabiei2010non} methods for joint phase and channel estimation. In our work, we will not focus on high computational methods or interpolation filters \cite{bello2023wiener,kuribayashi2022performance}, as they will not fit in the futuristic stringent latency requirements.


For SC-FDMA, phase noise manifests as fluctuating phase shifts between consecutive symbols, impacting time-domain signal integrity. The 5G-NR standard employs Pilot Reference Symbols (PTRS) for phase noise tracking and compensation \cite{etsi2022138}.

Studies, including \cite{sibel2018pilot}, have evaluated PTRS-based techniques within DFT-s-OFDM systems, highlighting two low-complexity compensation methods—common phase equalization (CPE) and linear interpolation (LI)—alongside more complex approaches like discrete cosine transform (DCT) and Kalman filters. Furthermore, \cite{khanzadi2013estimation} used pilots to find the maximum likelihood estimates of phase noise at the PTRS positions, followed by linear interpolation for data symbols. 
The literature distinguishes between Pre-DFT and Post-DFT PTRS placements \cite{Syrjala,Zheng}, with Pre-DFT, where reference symbols are time-domain pulses, being the choice for 5G-NR, facilitating phase offset estimation and subsequent interpolation for data symbol phase estimation.

Block PTRS allocation, initially proposed for the OFDM scheme \cite{mikko1,mikko2}, introduces PTRS as subcarriers, which means that at the receiver, it will capture the ICI effect introduced by the phase noise process. Then, it can be used to resolve the ICI in the data subcarriers of the OFDM symbol. Our investigation seeks a suitable adaptation for DFT-s-OFDM, proposing a Pre-DFT modulated block PTRS approach.

In this paper, we claim that the current PTRS allocation adopted by the 5G-NR standard is less than ideal for effective time-frequency analysis. The primary concern is that PTRS pulses exhibit bandwidth exceeding their sampling Nyquist rate, leading to potential aliasing issues in the phase noise sampling process. This paper advocates for a reevaluation of PTRS allocation to enhance phase noise compensation efficacy in DFT-s-OFDM systems.

In this paper, we draw an analogy with the Discrete Wavelet Transform's (DWT) multiresolution analysis to enhance PTRS allocation, aiming for a more refined sampling of the phase noise spectrum. By strategically allocating PTRS in a manner similar to the selective use of wavelets in DWT, we aim to reduce aliasing effects and improve phase noise estimation accuracy. Additionally, we introduce a block PTRS scheme similar to using lower-scale wavelets by employing pre-DFT block allocation. To support this allocation strategy, we also present corresponding modulator and demodulator schemes.

The remainder of the paper is organized as follows: Section II provides the DWT analogy with the two methods of PTRS allocation. Section III introduces the block PTRS allocation, illustrates the PTRS pulse shaping, and exemplifies phase noise sampling techniques. Section IV  details the proposed modulator and demodulator schemes. 
Section VI presents simulation results, validating the effectiveness of the proposed methodologies and algorithms.

\section{Pulse Allocation Analysis  for Phase Noise Sampling}
This section flows as follows: First, we introduce the background of phase noise sampling via PTRS and its time-frequency domain representation. Then, we discuss the aliasing effect resulting from sparse sampling and draw an analogy with DWT sampling.  

\subsection{Background}
In 5G-NR DFT-s-OFDM, PTRS symbols are allocated pre-transform precoding, or in other words, before the DFT block. This is equivalent to assigning DFT-s-OFDM pulses (sinc-shaped pulses) as PTRS between the data symbols to measure the phase noise at the position of the PTRS pulses. The formulation of the DFT-s-OFDM symbol $S$ is as follows.
\begin{equation}
\begin{split}
    S &= \mathbb{F}_n^{\dag} \Big[\mathbb{F}_m  \big[
        d_0,~d_1,\cdots,~d_{g},~p_0, \\
        &\quad d_{g+1},\cdots,~d_{2g},~p_1,~d_{2g+1},\cdots
        \big] \Big],
\end{split}
\end{equation}
where $d$ represents the digital modulated data symbols, $g$ denotes the number of data symbols between each two PTRS symbols $P$, with no PTRS symbols grouped in this formulation. $\mathbb{F}_m$ and $\mathbb{F}_n^{\dag}$ represent the operations: FFT of length m and IFFT of length n, respectively  ($n>m$).  

Notably, symbol $S$ manifests as a sequence of orthogonal sinc pulses, 
with the pulses adopting a circular shape at the IFFT output's boundaries.  DFT-s-OFDM is thus deconstructable into two symbol trains: data symbols and PTRS symbols, $S = S_{data} + S_{PTRS}$. The components $S_{data}$ and $S_{PTRS}$ are further detailed below, where $\mathbf{z}(g)$ signifies a zero sequence of length $g$. The data component is represented by:
\begin{equation}
\begin{split}
    S_{\text{data}} &= \mathbb{F}_n^{\dag} \Big[\mathbb{F}_m  \big[
        d_0,~\cdots,~d_{g},~0,~d_{g+1},\cdots, \\
        &\quad ~d_{2g},~0,~d_{2g+1},\cdots,~d_{3g},~0,~d_{3g+1}~\cdots   \big] \Big],
\end{split}
\end{equation}
and the PTRS component by:
\begin{equation}
    S_{\text{PTRS}} = \mathbb{F}_n^{\dag} \Big[ \mathbb{F}_m  \big[    \mathbf{z}(g),~p_0,~\mathbf{z}(g),p_1,~\mathbf{z}(g),~p_2,~\cdots   \big] \Big].
\end{equation}

 \begin{figure*}[b]
     \centering
         \includegraphics[width=0.9\textwidth]{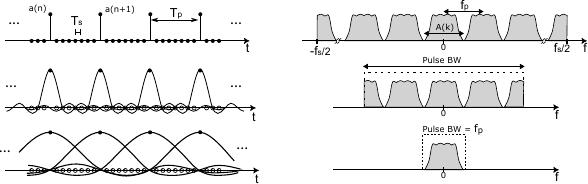}
         \caption{Revisiting Sampling Theory: We show three cases of sampling, the first one using Delta Dirac functions with rate $f_p$, the second is using sinc pulses with zero samples in between the sampling pulses, and the third case uses interleaved orthogonal sinc functions with width spanning the duration between the sampling instances. The pulses are not added in the illustration, and the signal value at the instance denoted with `o' is the sum of all the pulses at that instance.  }
         \label{fig:revisit_theory}
\end{figure*}
Our analysis predominantly focuses on the $S_{PTRS}$ component's capability to sample phase noises originating from both the transmitter ($\theta_{tx}$) and the receiver ($\theta_{rx}$). The time-domain received PTRS sequence is given by:
\begin{equation}
    R_{\text{PTRS}} = \left[ \big[S_{\text{PTRS}}.~e^{(j \theta_{tx})} \big] \otimes h(\tau) \right].~ e^{(j \theta_{rx})} + \mathfrak{w},
\end{equation}

where $h(\tau)$ is the channel impulse response with the delay $\tau$ matching the waveform sampling rate, $\otimes$ denotes the convolution operation, and $ \mathfrak{w}$ is the additive white Gaussian noise (AWGN). Simplifying for a single-tap channel (flat fading), we consolidate the phase noise effects into $\theta_{pn} = \theta_{tx} + \theta_{rx}$. Consequently, the processed PTRS vector post-DFT-s-OFDM demodulation is:
\begin{equation}
\begin{split}
    r_{PTRS} &= \underbrace{\mathbb{F}_m^{\dag} \big[~\overbrace{\mathbb{F}_n [R_{\text{PTRS}}]}^{\text{Frequency}}~\big]}_{\text{Time}}\\ 
    & = \mathbb{F}_m^{\dag} \Big[ \underbrace{\mathbb{F}_n [S_{PTRS}]}_{\text{Frequency}} \otimes \underbrace{\mathbb{F}_n \big[ \overbrace{{e^{(j\theta_{pn})}}}^{\text{time}} \big] }_{\text{Frequency}} + \mathbb{F}_n[\mathfrak{w}]   \Big]   \\
        &= \underbrace{ \mathbb{F}_m^{\dag} \mathbb{F}_n \left[S_{PTRS}\right]}_{\text{Time}} ~.~ \underbrace{\mathbb{F}_m^{\dag} \mathbb{F}_n \left[e^{(j\theta_{pn})}\right]}_{\text{Time}}  + \mathbb{F}_m^{\dag} \mathbb{F}_n[\mathfrak{w}].
\end{split}
\end{equation}

The first term of the above term corresponds to the transmitted PTRS sequence ($[  \mathbf{z}(g),~p_0,~\mathbf{z}(g),p_1,~\mathbf{z}(g),~p_2,~\cdots ]$), and the second term is the lowpass filtered phase noise process, with a rectangular shaped filter of length $m$. This means that the phase noise information is available at the PTRS positions but captured with the whole bandwidth of the PTRS and data symbols. This configuration results in phase noise being filtered through a bandwidth equivalent to $m$ bins at the PTRS locations$~k$. 
\begin{equation}
    r[k] = [\hat p_0,~\hat p_1,~\hat p_2,~\cdots]\cdot e^{(j\theta_{pn}[k])} + \mathfrak{w}[k]
\end{equation}

\begin{equation}
    \angle r[k] = [\angle \hat p_0,~\angle\hat p_1,~\cdots] +{\theta}_{pn}[k]  +  tan^{-1}(\frac{w_\perp}{A_{p}}),
\end{equation}

where $w_\perp$ is the Gaussian noise value projected on the axis perpendicular to the PTRS symbol vector, $A_{p}$ is the amplitude of the PTRS symbol.
The distribution  $P(w_\perp) 
= \int_{-\infty}^{\infty} 
   \frac{1}{2 \pi \sigma^2}  
    \exp{\left(-\frac{w_\perp^2 + w_\parallel^2}{2 \sigma^2} \right)} dw_\parallel 
   ~=~ \mathcal{N}(0,\sigma^2),$
   is the normal distribution with the same Gaussian noise standard deviation. At high SNR, we can use the approximation $ tan^{-1}(\frac{w_\perp}{A_{p}}) \approx \frac{w_\perp}{A_{p}}$.   
The estimated PN shifts is determined with reference to the known PTRS sequence:
\begin{equation}
    \hat{\theta}_{pn}[k] =\angle r[k] -[\angle p_0,~\angle p_1,~\angle p_2,~\cdots]
    \label{eq:deconv_sunbt}
\end{equation}
This means that this method of sampling will aggregate all the channel effects, including additive white Gaussian noise, with a bandwidth corresponding to $m$ bins. This adds to the aliasing problem in the case of a wide bandwidth sampling pulse, as discussed in the next subsection. 
In the next subsection, we mention the issue of this sparse sampling.

\subsection{Aliasing of Sparse Samples}
In this section, we summarize the sampling theory concepts and formulation of sampling pulses in time and frequency domains. This will give us insight into the representation of phase noise sampling and aliasing conditions. We use Fig. \ref{fig:revisit_theory} to follow up with the equations next. First, there is the well-known case of sampling with a train of Dirac Delta functions at a rate of $f_p$ and discretized time with a rate $f_s$. The time and frequency domain representations are
\begin{equation}
    x(t)= \sum_{n=-\infty}^\infty \delta(t- n T_p),
\end{equation}
\begin{equation}
    X(f)= \sum_{n=-\infty}^\infty \int_{-\infty}^\infty \delta(t- n T_p)~e^{-j2\pi f t} dt.
\end{equation}
The Fourier transform of a Dirac comb, also known as Shah-function, is another Dirac comb. 
\begin{equation}
    X(f)= f_p \sum_{k=-\infty}^\infty  \delta(f- k f_p).
\end{equation}
In the case of the Dirac pulses in the time domain are not a sequence of the same value but modulated by values $a(n)$, then the frequency domain representation becomes
\begin{equation}
    X(f)= f_p \sum_{k=-\infty}^\infty \delta(f- k f_p) \otimes A(f)=  f_p \sum_{k=-\infty}^\infty  A (f- k f_p),
    \label{eq:inftyfreqshah}
\end{equation}
where $A$ is the frequency domain representation of the $a$ sequence. 
When time is quantized with rate $f_s$, as in Fig. 1,  $X(f)$ is defined over the range $[-\tfrac{f_s}{2},~\tfrac{f_s}{2}]$, and the integer $|k|< \tfrac{f_s}{2 f_p}$.

Now, we move to the case when the sampling impulse has a shape, particularly a sinc shape because it fits with the studied system (DFT-s-OFDM). The train of sinc functions with bandwidth $B$ at intervals $T_p$ is formulated as:
\begin{equation}
    x(t)= \sum_{n=-\infty}^\infty a_n sinc(B( t- n T_p)),
\end{equation}
where $sinc(B t)= \frac{sin(\pi B t)}{(\pi B t)}$.
The Fourier transform gives the frequency domain representation
\begin{equation}
    X(f)= \int_{-\infty}^\infty {\textls[0]{III}}_{T_p}(t) \otimes sinc(Bt))~e^{-j2\pi f t} dt,
\end{equation}
where ${\textls[0]{III}}_{T_p}(t)$ is the Dirac comb with period $T_p$. This corresponds to the multiplication of the frequency domain representation  $X_{\textls[0]{III}}(f)$ in Eq. \ref{eq:inftyfreqshah} with the transform of a sinc function
\begin{equation}
    X(f)= X_{\textls[0]{III}}(f) X_{sinc}(f),
\end{equation}
\begin{equation}
    X(f)= \frac{1}{T_p} rect(\frac{f}{B}) \sum_{k=-\infty}^\infty  A(f- k f_p),
\end{equation}
$rect$ function is the inverse of $sinc$ function, and it is non-zero only for $|f| < \frac{B}{2}$.

\begin{equation}
    X(f) = \frac{1}{T_s} \sum_{k} A_n(f - k f_s), \quad \forall k \in \mathbb{Z}, \ |k f_s| < \frac{B}{2}.
\end{equation}. 

To understand the progression between the second and third cases and why we present them in this way, we need to relate this to the PTRS DFT-s-OFDM sampling pulses. The pre-DFT PTRS allocation results in sinc-shaped sampling pulses with width to fit between data symbols and stay orthogonal. Hence, $B$ is the same as the bandwidth of the DFT-s-OFDM symbol, which is the number of symbols times the subcarrier spacing $B=m~\Delta f$. 

Let us present the third case where the sampling sinc pulses do not have space between them for data pulses, and their width spans the maximum duration that maintains orthogonality. In that case, the bandwidth decreases to $B=1/T_p$. We can see the difference between the two frequency domain representations and realize that when this spectrum is convolved with the phase noise spectrum (corresponding to sampling in the time domain), we will have potential aliasing for the second case. The convolution with the spectrum of the PTRS sequence ($A(f)$) is resolved at the receiver at the step of division by the complex values of the symbols in the time domain, leading to the phase subtraction in Eq. \ref{eq:deconv_sunbt}, and deconvolution with $A(f)$ in the frequency domain. 



\begin{figure}[h]
    \centering
   \begin{subfigure}[b]{0.48\textwidth}
         \centering
         \includegraphics[width=\textwidth]{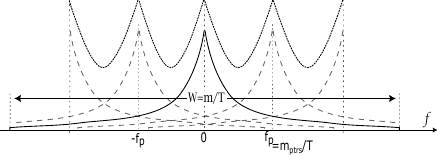}
         \caption{}
         \label{fig:aliasing_a}
     \end{subfigure}
     \begin{subfigure}[b]{0.48\textwidth}
     \hspace{-1pt}
         \includegraphics[width=\textwidth]{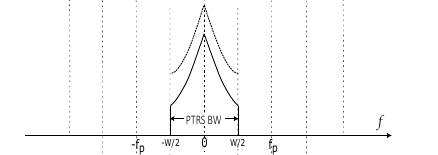}
         \caption{}
         \label{fig:aliasing_b}
     \end{subfigure}
    \caption{Comparison between the interpolation results for the captured phase noise in two distinct PTRS pulse configurations. (a) Depicts the aliasing effect.
(b) Illustrates the corresponding filtered spectrum in comparison with the interpolated one.}
    \label{fig:aliasing}
\end{figure}

Consider the scenario depicted in Fig. \ref{fig:aliasing}, where we have a phase noise process with a bandwidth higher than the PTRS rate, i.e., the phase noise varies faster than the allocated PTRS rate. We are comparing the sparse PTRS wide bandwidth pulses versus the low bandwidth compact PTRS pulses, both with the same repetition rate.

In Fig. \ref{fig:aliasing_a}, where a DFT-s-OFDM symbol comprises $m_{data}$ and $m_{ptrs}$ symbols, thus $m = m_{data} + m_{ptrs}$. With a uniform PTRS distribution, the repetition rate is $\frac{m_{ptrs}}{T}$, with $T$ denoting the DFT-s-OFDM symbol duration.  However, the PTRS frequency domain window is $\frac{m}{T}$, surpassing the repetition rate. This discrepancy leads to aliasing following the interpolation step.

Conversely, as depicted in Fig. \ref{fig:aliasing_b}, when PTRS is allocated to a distinct DFT-s-OFDM symbol without intervening data (as in the third case in Fig. \ref{fig:revisit_theory}), it affords the flexibility to tailor the sinc pulses' width and repetition rate. Assuming PTRS pulses share the repetition rate $\frac{m_{ptrs}}{T}$, and the sinc pulse's main lobe spans the entire $\frac{T}{m_{ptrs}}$ duration, the PTRS frequency domain window aligns with the repetition rate. This alignment ensures aliasing is circumvented, enabling the receiver chain to effectively filter the phase noise process and recover the phase noise process.

The relationship between the PTRS pulse repetition frequency and frequency domain window occupation is critical to avoid the aliasing effect. In the next section, we show examples of PTRS pulse allocations and widths to understand the relationships of the above parameters. Then, gradually, we make deductions about optimized pulse shaping that matches the repetition rate with the pulse window occupancy. 

\subsection{Drawing Parallels: DWT and Sparse Sampling}
\begin{figure}[h]
    \centering
    \includegraphics[width=0.49\textwidth]{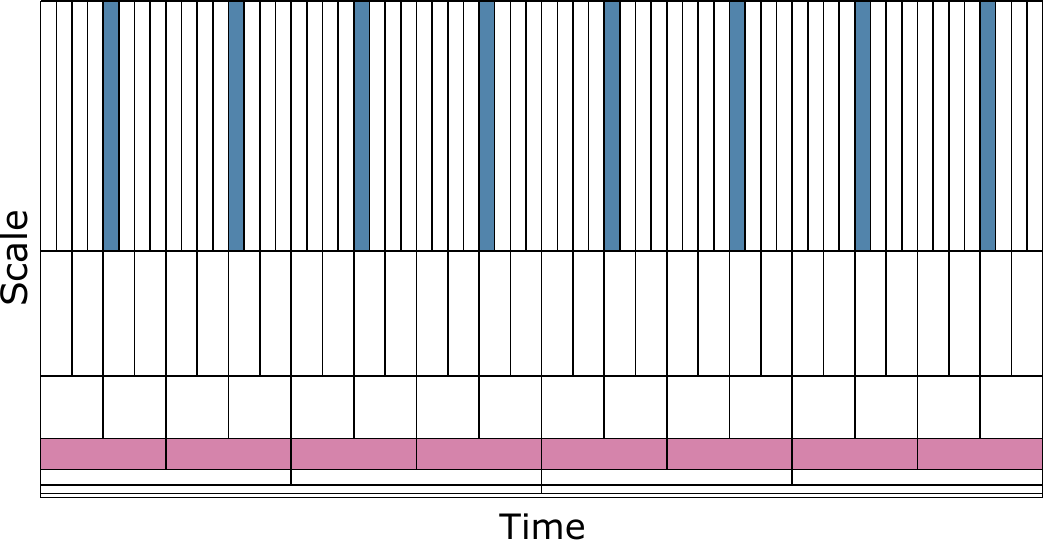}
    \caption{Dyadic Wavelet Transform. The shaded wavelets are analogous to the 5G-NR and the proposed block PTRS allocations.}
    \label{fig:dyadic}
\end{figure}

In contrast to the DWT, which uses all the wavelets for sampling, 5G-NR employs PTRS pulses sparsed by a number $g$ of slots (filled with data pulses). This approach essentially selects a sparse array of higher-scale sampling wavelets, as depicted in Figure \ref{fig:dyadic}. However, this strategy leads to a discrepancy where the sampling rate falls below the bandwidth of the pulses being sampled, consequently inducing aliasing. 

Instead, the proper practice is to use a bandwidth of the wavelet or pulse that matches the sampling rate (the repetition rate of the PTRS pulses). This strategy is exemplified by lower scale sinc wavelets or pulses, as illustrated in Fig. \ref{fig:dyadic}. 

In addressing phase noise within DFT-s-OFDM systems, this work proposes a PTRS allocation strategy inspired by principles of the DWT. The DWT, renowned for its efficient time-frequency localization, offers a novel perspective on optimizing phase noise sampling and tracking.
In this section, we explain the 5G-NR PTRS allocation formula and demonstrate its equivalence to sampling with a sinc-shaped pulse. Then, we map it to a discrete wavelet transform with a sinc-mother wavelet.  

DWT is an essential tool in time-frequency signal analysis,
enabling effective signal representation and compression. In dyadic DWT, wavelets undergo discrete sampling, with the mother wavelet, denoted as $\Psi_{j,k}$, subjected
to shifts and scaling by powers of two, as follows:  
\begin{equation}
    \Psi_{j,k}=\dfrac{1}{\sqrt{2^j}} \Psi \Big( \dfrac{t-k2^j}{2^j} \Big),
\end{equation}
where $j$ is represents the scale parameter and $k$ the shift parameter, both integers. This operation generates various tilings with different time-frequency resolutions, illustrated in Figure \ref{fig:dyadic}. The wavelet coefficient $\gamma_{j,k}$ of a signal $x(t)$ is the projection of $x(t)$ onto the wavelet:
\begin{equation}
    \gamma_{j,k} = \int_{-\infty}^{\infty} x(t) \dfrac{1}{\sqrt{2^j}} \Psi \Big( \dfrac{t-k2^j}{2^j} \Big) ~dt .
\end{equation}
A mother wavelet can take on various forms. However, our analysis considers using sinc-shaped mother wavelets to create orthonormal DFT-s-OFDM-like pulses, as discussed above. 


\section{Block PTRS Realization}
This section examines PTRS allocation schemes, presuming distinct streams for data and PTRS symbols, each modulated onto separate DFT-s-OFDM symbols. We primarily focus on the modulation of PTRS symbols without interspersed data, aiming to identify optimal pulse allocations. The ensuing section will depict the integrated transmission of DFT-s-OFDM PTRS and data symbols.

\begin{figure}[h]
    \centering
    \includegraphics[width=0.48\textwidth]{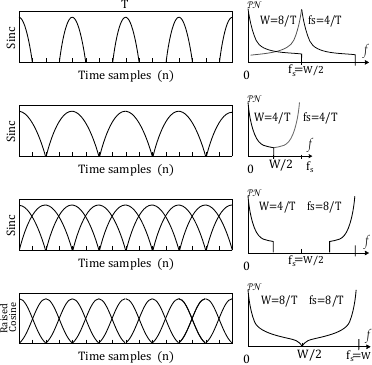}
    \caption{Block PTRS allocation examples.}
    \label{fig:aloc}
\end{figure}

In Fig. \ref{fig:aloc}, different cases of PTRS allocations are presented. The figure shows only the main lobes of the pulses in the time domain representations on the left plots, where the main lobe of a sinc pulse is the region between the first zeros around the pulse center. The time duration of the DFT-s-OFDM symbol is $T$, the DFT block size $m$ varies with the number of allocated PTRS pulses, and the IFFT size is $n$, ($n >> m$). The pulse oversampling rate is $n/m$.

On the right side of Fig. \ref{fig:aloc}, we show the phase noise spectrum representation with the sampling effect depicted in the replicated spectrum at the PTRS sampling rate $f_s$ (the repetition rate). Aliasing is caused by the relationship between frequency domain window W and sampling frequency fs. For each row, we change the pulse duration and the repetition rate and discuss the aliasing effect. 

In the first row plot, We have four sinc pulses with main lobe duration $T/8$. Hence, the repetition rate is $f_s=4/T$, and the window size is $W=8/T$. The symbol can be created through the operation $S_{PTRS}= \mathbb{F}_n^{\dag} \Big[ \mathbb{F}_{16} [ 1 ~ \mathbf{z}(3)~ 1~  \mathbf{z}(3)~ 1 ~\mathbf{z}(3) ~1~  \mathbf{z}(3) ]$. In a practical PTRS symbol, the ones are replaced with PTRS symbol values, e.g., QAM modulated values.  The resultant sampling rate to window size relationship is $W>f_s$, which leads to aliasing, as shown in the phase noise spectrum plot.

In the second row plot, the same pulse repetition rate is assigned $f_s=4/T$, but the Sinc pulses are wider in the time domain, with main lobe duration $T/4$. This corresponds to a window size of $W=4/T$. The symbol can be created through the operation $S_{PTRS}= \mathbb{F}_n^{\dag} \Big[ \mathbb{F}_{8} [ p[1] ~ 0 ~ p[2]~ 0 ~ p[3] ~0 ~p[4]~  0]$, where $p$ is the PTRS symbols.  The resultant sampling rate to window size relationship is $W=f_s$, which avoids aliasing, as shown in the phase noise spectrum plot. However, the PTRS stream has zero symbols, which is inefficient because they can be used to achieve a higher sampling rate. 
 
In the third row plot, we increase the repetition rate by utilizing the orthogonality of the sinc pulses. The symbol can be created through the operation $S_{PTRS}= \mathbb{F}_n^{\dag} \Big[ \mathbb{F}_{8} [ p[1] ~  p[2]~  p[3] ~p[4]~p[5] ~  p[6]~  p[7] ~p[8]~ ]$. The resultant sampling rate to window size relationship is $f_s=W/2$, which avoids aliasing and gives room to use higher bandwidth pulses. Therefore, in the fourth-row plot, we show the example of using raised cosine pulses, the roll-off factor of unity. This pulse shape exhibits the same orthogonality properties as the sinc pulse, but it takes twice the bandwidth. Hence, it covers more bandwidth of the phase noise spectrum while still avoiding aliasing ($W=f_s$). 

Using a raised cosine pulse comes at the expense of using an FFT size double the size of the sinc pulse allocation to accommodate the windowing after the DFT block. This could have covered a sequence of sinc pulses (third plot) with twice the sampling (main lobe=$T/16$, $fs=16/T$). In the simulations section, we compare those two modes to see if the usage of the raised cosine pulse is worth sacrificing for the PTRS sampling rate. In the next section, we propose a transceiver design for the above block PTRS allocation.

\section{Pre-DFT Block PTRS Allocation: Transceiver Design}
\label{trxsec}

\begin{figure}
    \centering
   \begin{subfigure}[b]{0.38\textwidth}
         \centering
         \includegraphics[width=\textwidth]{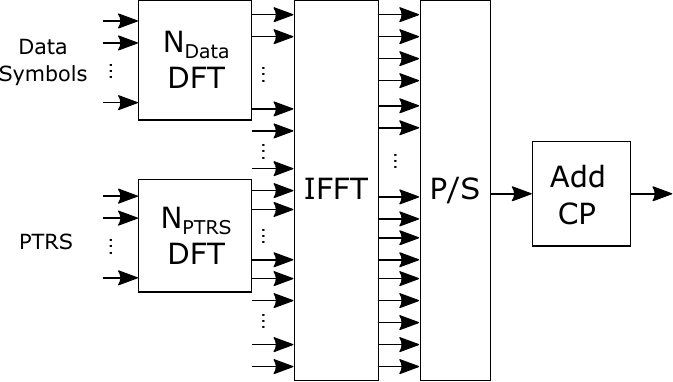}
         \caption{}
         \label{fig:blocka}
     \end{subfigure}
     \begin{subfigure}[b]{0.48\textwidth}
         \centering
         \includegraphics[width=\textwidth]{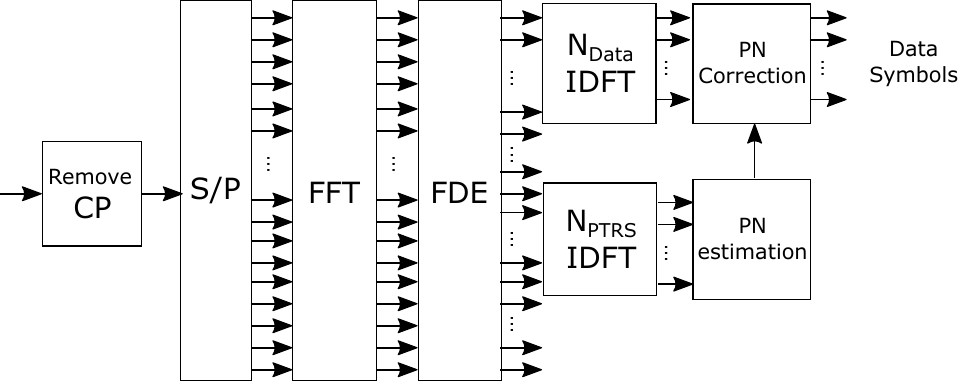}
         \caption{}
         \label{fig:blockb}
     \end{subfigure}
    \caption{Block PTRS allocation transceiver on a time-frequency resource grid.  }
    \label{fig:blocktrx}
\end{figure}
The implementation of block PTRS allocation is encapsulated within the modulator and demodulator setup, as depicted in Fig. \ref{fig:blocka}. This design facilitates the generation of two distinct streams for data and PTRS symbols, each modulated onto a separate DFT-s-OFDM symbol. Specifically, data symbols undergo transformation via a DFT block of length $m_{data}$, while PTRS symbols are processed through a DFT block of length $m_{PTRS}$. The outputs from these blocks are then integrated into the appropriate spectral bands within the IFFT block, following the modulation equation:
\begin{equation}
    Tx =\mathbb{F}_N^{\dag}\big[ \mathbb{F}_{m_{data}} [d] ~\Vert ~\mathbf{z}( g_f)~ \Vert~ \mathbb{F}_{m_{PTRS}} [p] \big],
\end{equation}
where $\Vert$ denotes concatenation, $d$ is the I/Q modulated data vector, $p$ is the I/Q modulated PTRS vector, and $g_f$ is a frequency-domain gap between the two bands, with $N>m_{data} > m_{PTRS}$.

\begin{figure}[h]
    \centering
    \includegraphics[width=0.45\textwidth]{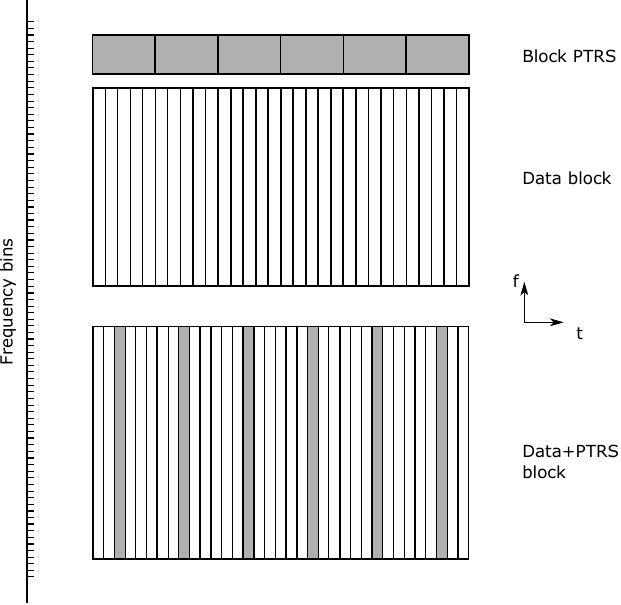}
    \caption{Block PTRS allocation in comparison with 5G-NR allocation.}
    \label{fig:blckallocation}
\end{figure}

The representation of the block PTRS modulation in the time-frequency grid is shown in Fig. \ref{fig:blckallocation}. The figure contrasts the block PTRS allocation with the conventional 5G-NR strategy, highlighting the temporal and spectral implications of each. 
The number of PTRS and data symbols are the same in both cases. Note that the time-frequency occupancy for the total data and PTRS is the same for both configurations. 
As the block PTRS configuration has the same duration as the 5G-NR configuration, the modulated pulses in the 5G-NR will be narrower in time and wider in frequency, taking the same frequency bins of the data and PTRS combined in the block configuration. 

The vertical strips seen in each block symbolize either PTRS or data symbols. Being modulated via DFT-s-OFDM, these are visualized as narrow pulses in time, spanning the designated spectral bands. An illustrative case within the figure features six PTRS symbols allocated to a block of six subcarriers post-DFT operation, dispersing the PTRS pulses across the block's duration. Conversely, in the standard configuration, PTRS spans the entire spectrum, leading to a more compressed time-domain representation. Independent of the chosen allocation method, phase shifts at PTRS positions are estimated and interpolated for subsequent analysis.

\begin{figure}[h]
    \centering
   \begin{subfigure}[b]{0.48\textwidth}
         \centering
         \includegraphics[width=\textwidth]{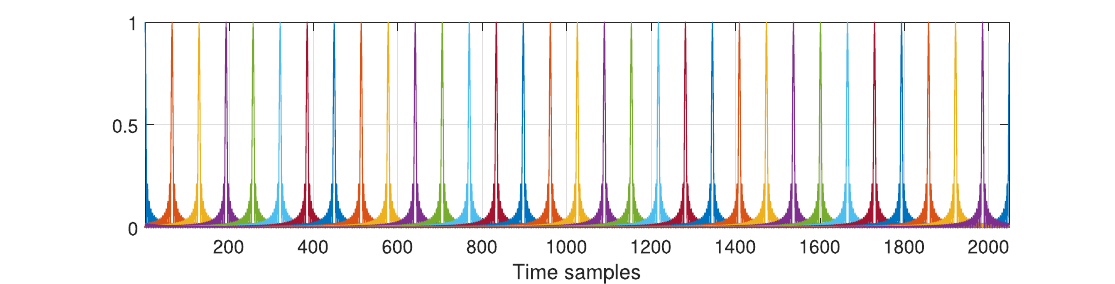}
         \caption{}
         \label{fig:ptrspulsea}
     \end{subfigure}
     \begin{subfigure}[b]{0.48\textwidth}
         \centering
         \includegraphics[width=\textwidth]{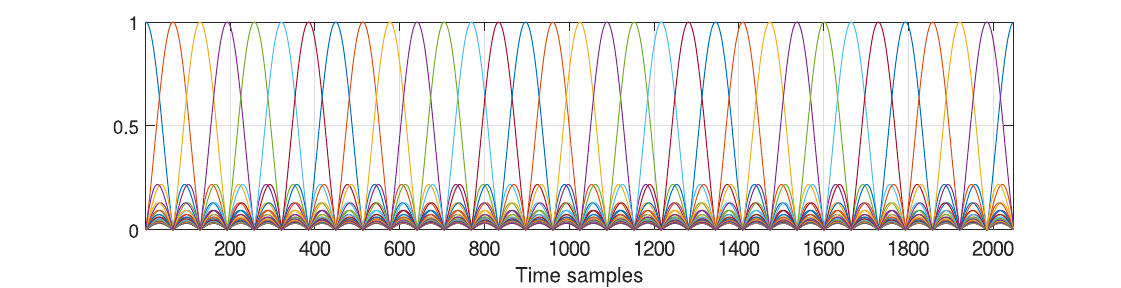}
         \caption{}
         \label{fig:ptrspulseb}
     \end{subfigure}
     \begin{subfigure}[b]{0.48\textwidth}
         \centering
         \includegraphics[width=\textwidth]{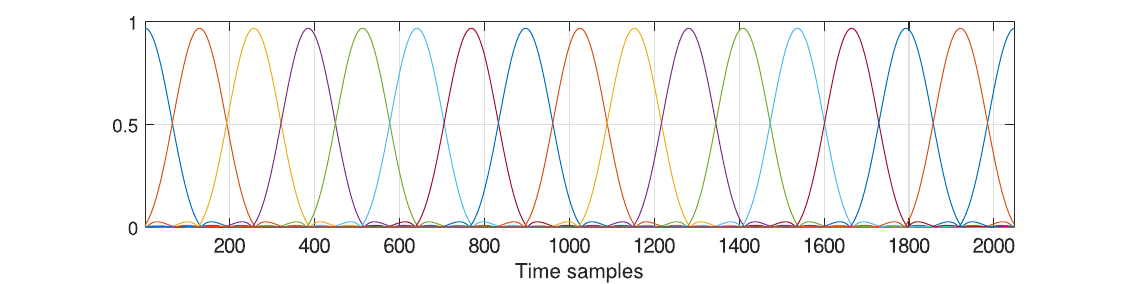}
         \caption{}
         \label{fig:ptrspulsec}
     \end{subfigure}
    \caption{PTRS pulses in time-domain (pulses are plotted individually).  }
    \label{fig:ptrspulses}
\end{figure}

\begin{figure*}[hb]
    \centering
   \begin{subfigure}[b]{0.87\textwidth}
         \centering
         \includegraphics[width=\textwidth]{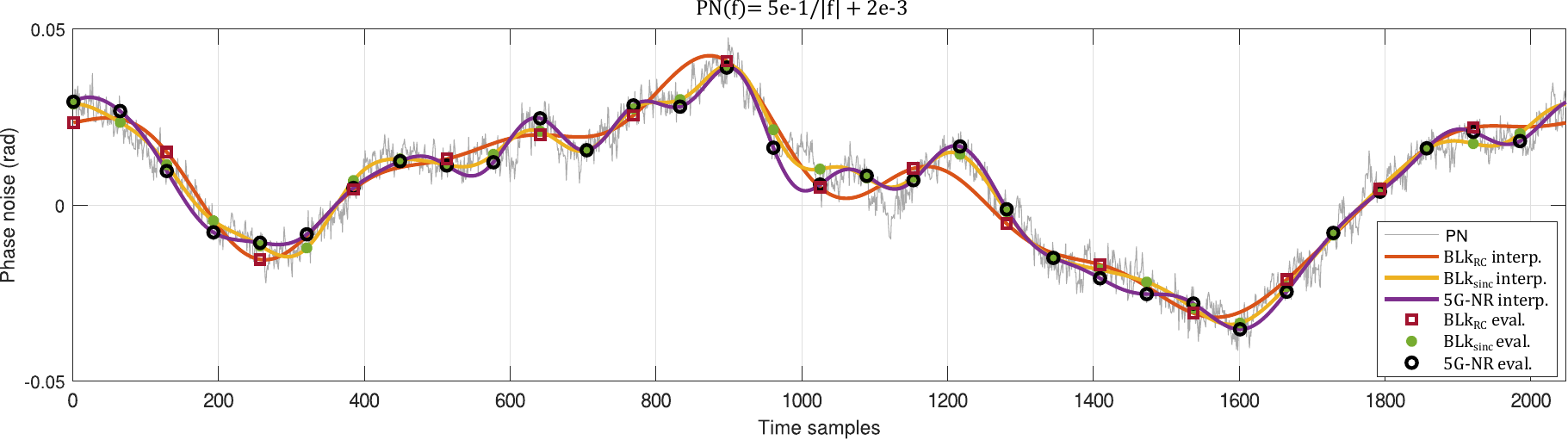}
         \caption{}
         \label{fig:trackinga}
     \end{subfigure}
     \begin{subfigure}[b]{0.87\textwidth}
         \centering
         \includegraphics[width=\textwidth]{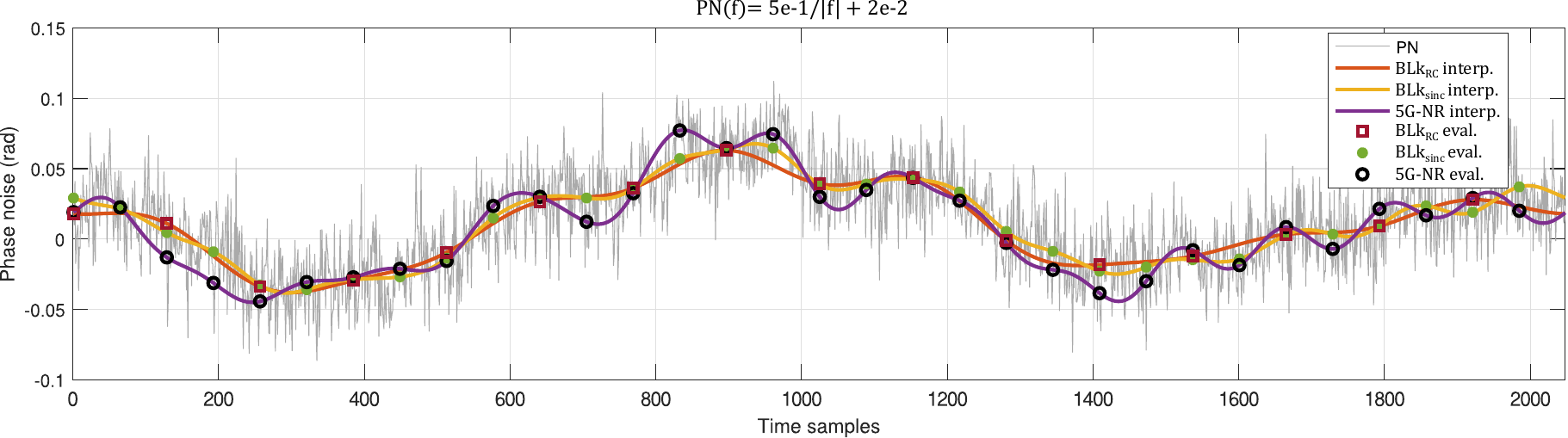}
         \caption{}
         \label{fig:trackingb}
     \end{subfigure}
     \begin{subfigure}[b]{0.87\textwidth}
         \centering
         \includegraphics[width=\textwidth]{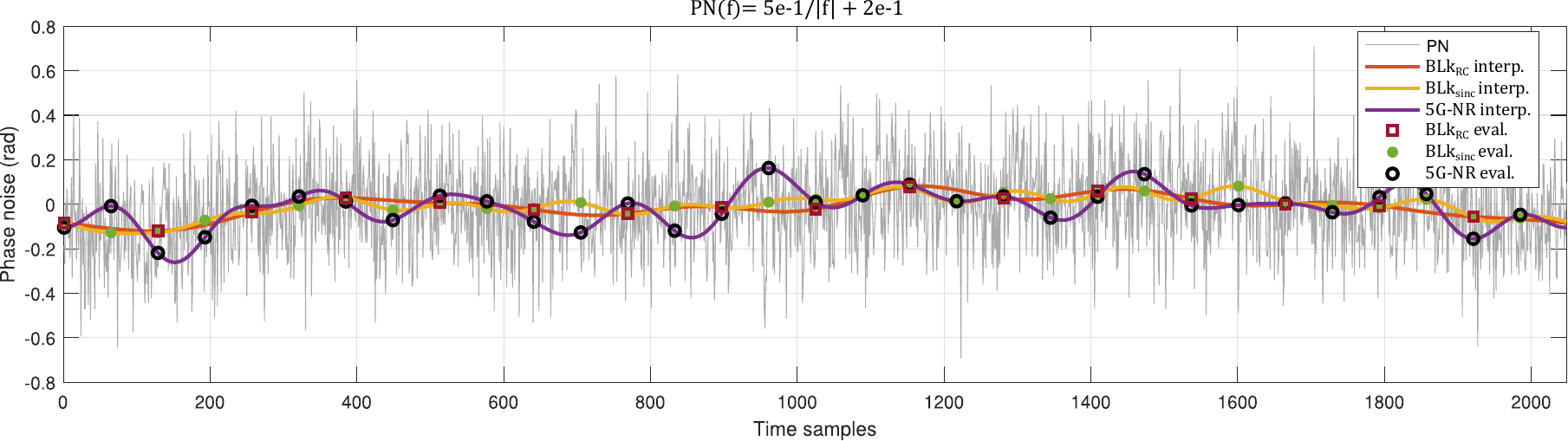}
         \caption{}
         \label{fig:trackingc}
     \end{subfigure}
    \caption{Phase noise tracking performance for different PTRS allocations and phase noise spectrums.  }
    \label{fig:tracking}
\end{figure*}

Gathering the PTRS information in a single band that does not spread over the full data bandwidth has several impacts. For instance, the PTRS band is narrower in the frequency domain. Therefore, the effect of a deep frequency fade or frequency null is less probable. Also, in the presence of an adaptive system, the PTRS band can be dynamically allocated to favorable bands without being tied down to the harsh requirements of the faster data stream. Moreover, the literature has shown that the frequency-selective channel effect with phase noise effect results in ISI for DFT-s-OFDM symbols, which degrades the phase noise estimations of the PTRS. In our case, we assign a narrower band to the PTRS, hence less channel selectivity and ISI effect, further enhancing the estimation results.

As previously noted, sampling a wideband PTRS pulse typically yields an estimate reflecting the broad spectrum of the phase noise process. The constrained sampling rate, mirroring the PTRS rate, predisposes this estimation to aliasing, potentially skewing the accuracy of lower-frequency component predictions. Conversely, sampling from a narrowband block PTRS generates a more refined estimate, indicative of a filtered phase noise profile. Aliasing in the context of wideband phase noise sampling may inadvertently amplify error-contributing higher-frequency components.

At the receiver, as illustrated in Fig. \ref{fig:blockb}, a frequency domain equalization (FDE) is applied after the FFT block to simultaneously equalize the data and PTRS bands. The data and PTRS frequency bins are then transformed back to their time domain shape using IFFT. A phase noise estimation block then constructs the estimated phase shift vector, which is utilized by a correction block to adjust the phase of the data stream accordingly. 

\section{Simulations and Results}



In this section, we compare the performance of the conventional 5G-NR and block PTRS allocations within different phase noise spectrums. The evaluation parameter is the root mean square phase error $\varepsilon_{rms}$ at the data indices $n_d$;
\begin{equation}
    \varepsilon_{rms}= \sum \sqrt{(\theta_{PN}-\hat{\theta}[n_d])^2} / m_{data},
\end{equation}
and the estimated phase noise $\hat{\theta}$ is found by referring the transmitted PTRS symbol to the received ones and then interpolating along the data indices. The interpolation used in this section is a sinc interpolator, formulated as:
\begin{equation}
\hat{\theta}= \mathbb{F}^{\dag}  [ W. \mathbb{F} [\hat{\theta}[P_0] ~\mathbf{z}[g]~\hat{\theta}[P_1] ~\mathbf{z}[g]~\cdots ]],
\end{equation}
where $W$ is a rectangular window low pass filter, capturing the sampling (PTRS repetition) rate. Note that in our evaluations, the channel between the transmitter and the receiver is assumed to be a single-tap channel, and we avoid channel equalization effects on performance. Moreover, we do not get bit-error-rate (BER) or block error rate (BLER) calculations, as all of the above will be dependent on extra assumptions of the channel type, symbol digital modulation, channel equalization type, and error coding scheme. We try to get a generic evaluation only based on phase error with different phase noise spectrums. 

\begin{figure}[h]
     \centering
     \includegraphics[width=.45\textwidth]{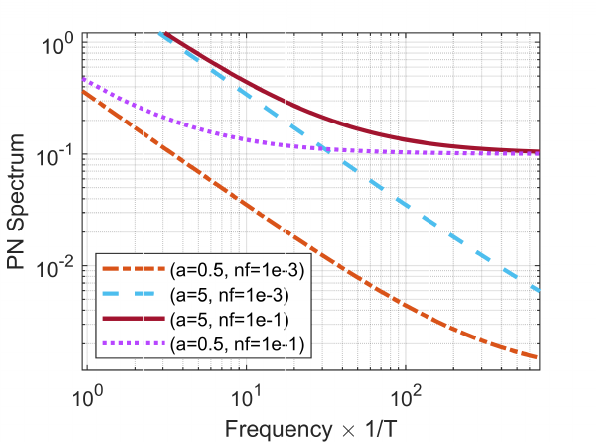}
     \caption{Monte Carlo simulation boundaries of amplitude and noise-floor parameters.}
     \label{fig:corners}
 \end{figure}

 \begin{figure}[h]
     \centering
     \includegraphics[width=.48\textwidth]{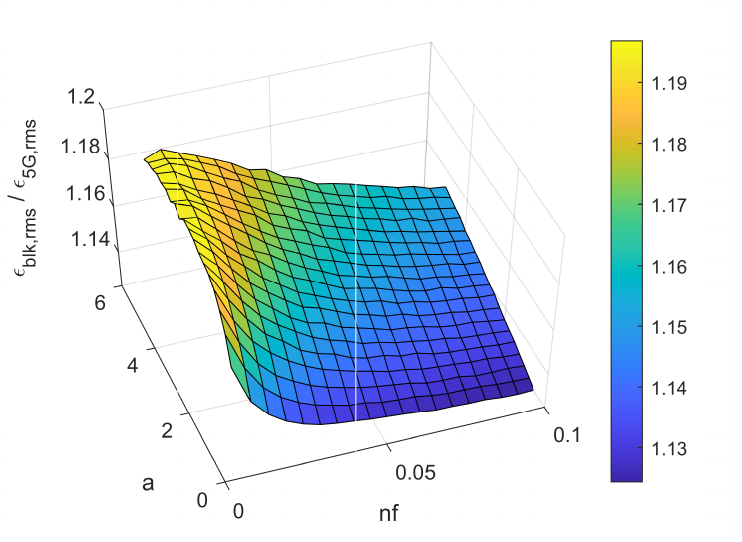}
     \caption{Block PTRS allocation gain over 5G-NR.}
     \label{fig:blkgain}
 \end{figure}
 \begin{figure}[h]
     \centering
     \includegraphics[width=.48\textwidth]{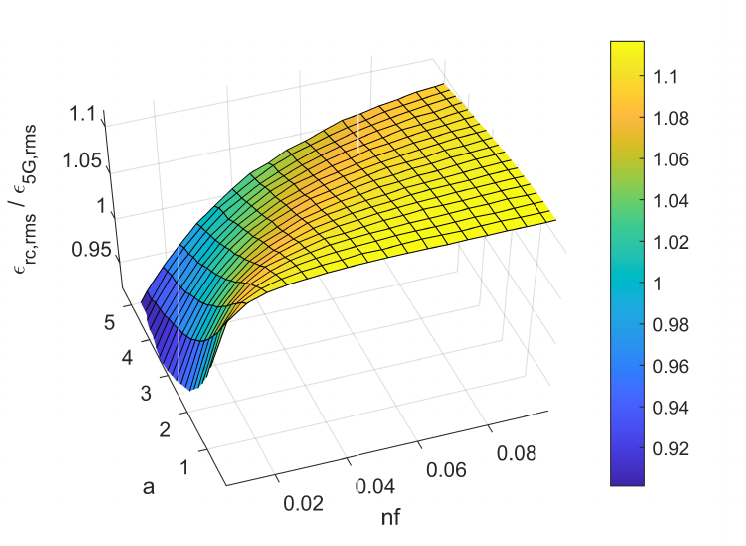}
     \caption{Raised Cosine Block PTRS allocation gain over 5G-NR.}
     \label{fig:rcgain}
 \end{figure}

\begin{figure*}[ht]
         \centering
         \includegraphics[width=\textwidth]{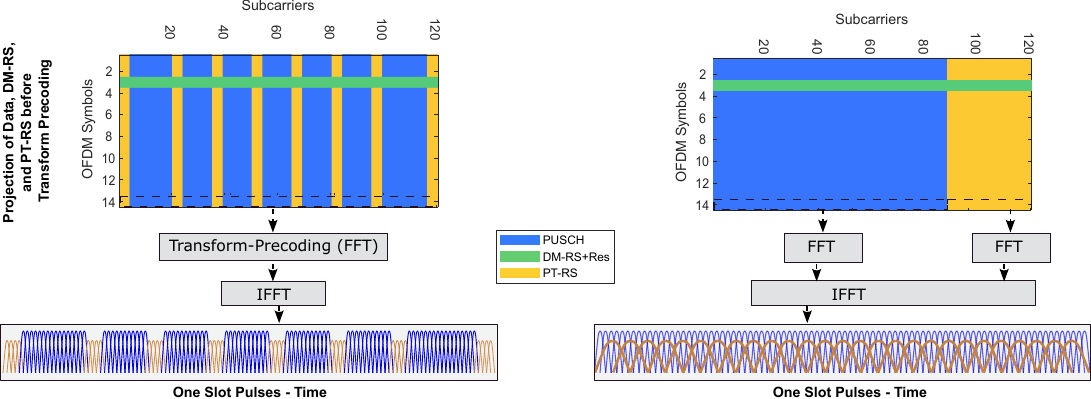}
    \caption{Comparison of the 5G NR PTRS allocation ($[N_{samp}^{group},~ N_{group}^{PT-RS}] = [4,~ 8]$), with the proposed block-based PTRS scheme. Each slot consists of 14 symbols, and the figure highlights the pulses of the final symbol. Note that we only illustrate the PTRS pulses' main lobe. In the block-based scheme, the pulses are plotted in absolute magnitude, omitting the higher-band modulation details for clarity.   }
    \label{fig:5GcompPlan}
\end{figure*}

\subsection{Three Different Spectrum Demonstration}

Shown in Fig. \ref{fig:ptrspulses}, the pulse shapes in the PTRS symbols for the different allocations under study. For demonstration, the pulses are overlayed on top of each other with different colors and not added.
The number of PTRS symbols is 32 for the 5G-NR and block allocations, with sinc-shaped pulses, and 16 for the raised cosine-shaped pulses. The pulses are distributed over an IFFT size of 2048. We assume $(128-32)=96 $ data symbols in the case of 5G-NR allocation. And 128 data symbol in the case of block allocations because it is a full DFT-s-OFDM symbol transmitted in Parallel with the PTRS symbol, as discussed in Section \ref{trxsec}, in Fig. \ref{fig:blckallocation}.
The DFT size for the 5G-NR symbol is 128, while for the block PTRS case, the PTRS symbol and data symbol are 32 and 128, respectively.

We present the performance of the PTRS tracking in Fig. \ref{fig:tracking}.
The phase noise spectrums used for evaluation follow the distribution
\begin{equation}
    PN(f)=\frac{a}{\lvert f \rvert} + n_f ~~(dBc/Hz), 
\end{equation}
where $n_f$ is a noise floor for the phase noise process, and $a$ is an amplitude parameter. The plots in Fig. \ref{fig:tracking} share the same phase noise amplitude value $a=0.5$, and the noise floor values are $n_f=2\times 10^{-3}$ for  Fig. \ref{fig:trackinga},  $n_f=2\times 10^{-2}$ for  Fig. \ref{fig:trackingb}, and  $n_f=2\times 10^{-1}$ for  Fig. \ref{fig:trackingc}. The frequency $f$ is normalized to the frequency bin spacing ($\frac{1}{T}$). We observe that the 5G-NR allocation is less immune than the block schemes in the presence of higher harmonic variations of the phase noise random process. The block schemes have an averaging effect that follows the lower phase noise harmonics and filters out the higher phase noise harmonics. 

As the phase noise floor decreases, the lower phase noise harmonics become dominant. Hence, the tracking of the block and 5G-NR becomes more efficient. But still, the block scheme with a sinc shape performs better at the same repetition rate. The raised cosine-shaped PTRS, on the other hand, does not track properly due to its lower repetition rate. Therefore, we can deduce that trading the repetition rate for a wider bandwidth pulse is not worth it. Next, we sweep over the phase noise spectrum parameters. 

\subsection{Monte Carlo Simulation}
In this subsection, we sweep the two-phase noise spectrum parameters $a$ and $n_f$, and we evaluate the gain of the Block PTRS allocation method in comparison to the 5G-NR allocation. The ranges for the amplitude and noise-floor parameters are [0.5, 5] and [1e-1 1e-3], respectively. The phase noise spectrums corresponding to the boundaries of the range are plotted in Fig. \ref{fig:corners}.

 The rms phase errors of the simulated methods at these boundaries are calculated and depicted in Table 1. The gain values of the proposed scheme $G_{BLK}$ and its raised cosine version $G_{BlkRC}$ are included in the table, and formulated as: $G_{BLK}= {\varepsilon_{rmsBLK}}/{\varepsilon_{rms5GNR}}$, and $G_{BlkRC}= {\varepsilon_{rmsBlkRC}}/{\varepsilon_{rms5GNR}}$.
 \begin{table}[h]
\caption{RMS phase error results at the Monte Carlo simulation boundaries.}
\label{tab:monte}
\centering
\resizebox{\columnwidth}{!}{
\begin{tabular}{|l|c|c|c|c|c|}
\hline
   &   $\varepsilon_{rms_{Blk}}$  &  $\varepsilon_{rms_{RC}}$     &  $\varepsilon_{rms_{5GNR}}$   & $G_{Blk}$   &  $G_{RC}$ \\ \hline
nf=1e-3, a=0.5    & 0.0017    &   0.0024  &  0.0020  &  1.18  &   0.83    \\ \hline
nf=1e-3, a=5   & 0.0144    &   0.0196   &    0.0170 &   1.18 &   0.87    \\ \hline
nf=1e-1, a=5    &  0.0521   &   0.0549    &  0.0601  & 1.15   &    1.09   \\ \hline
nf=1e-1, a=0.5    &  0.0446   &   0.0449   &  0.0502   &  1.13  &  1.12     \\ \hline
\end{tabular}}
\end{table}

The gain values across the noise spectrum parameter ranges are plotted in Fig. \ref{fig:blkgain} for the block PTRS and Fig. \ref{fig:rcgain} for the raised cosine version of the block PTRS. We realize that the block PTRS with sinc-shaped pulses is performing better across the full range but with more gain at the lower $nf$ and higher $a$ values. This is in the region where the phase noise error is low. Hence, we will perform better at high QAM modulations.

On the other hand, the raised cosine-shaped block PTRS fails to be better at lower $nf$ because of its lower repetition rate. But, as the $nf$ goes higher and $a$ goes lower, the RC block PTRS becomes better because of its averaging effect. Moreover, at its highest performance, the RC block does not perform better than the sinc block PTRS. We can rule out the idea of using an RC pulse with $\alpha=1$ in the context of PTRS sampling because it sacrifices the repetition rate to achieve the same bandwidth. 

 \subsection{Comparison with 5G NR}
In this section, we compare the performance of the proposed block-based PTRS scheme with the densest PTRS configuration specified by the 5G NR standard for DFT-s-OFDM. Specifically, for the 5G NR configuration, we employ the maximum-density mode ($[N_{samp}^{group},~ N_{group}^{PT-RS}] = [4,~ 8]$), where eight PTRS groups span the allocated bandwidth, each containing four PTRS. This arrangement yields the highest allowable PTRS density in uplink DFT-s-OFDM, providing a rigorous benchmark against our proposed block-based method, which uses 32 contiguous PTRS symbols, as illustrated in Fig. \ref{fig:5GcompPlan}.
We employ the Ericsson phase noise model \cite{R1-2003851} covering center frequencies of 10, 20, 50, and 100 GHz—to capture realistic phase noise effects in the millimeter-wave range. The key simulation parameters are summarized in Table \ref{tab:sim_params}.

\begin{figure}[h]
     \centering
     \includegraphics[width=.47\textwidth]{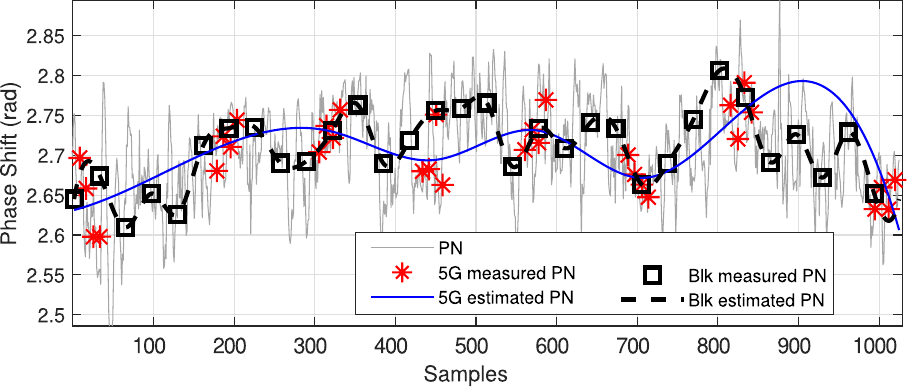}
     \caption{Phase-noise tracking at 10 GHz for the 5G NR and proposed block-based PTRS schemes. The block-based method more closely follows the noise variation than the 5G NR, using spline interpolation in both cases.}
     \label{fig:trackingsnapshot}
 \end{figure}

 \begin{figure}[h]
    \centering
    \begin{subfigure}[b]{0.44\textwidth}
     \hspace{-10pt}
         \includegraphics[width=\textwidth]{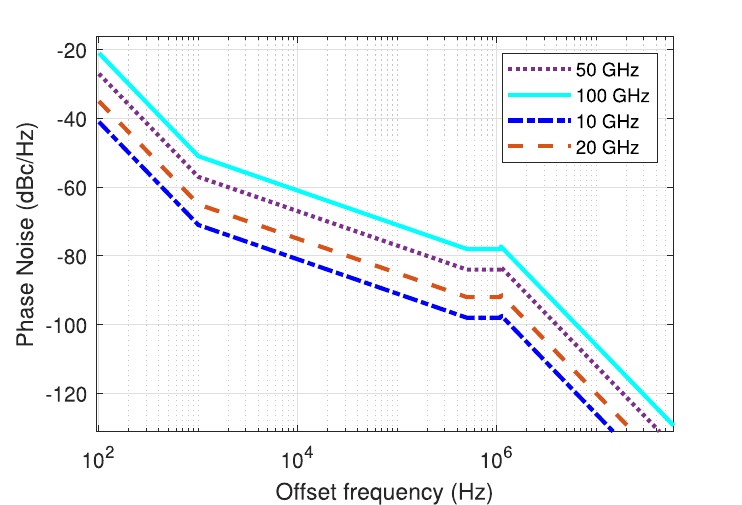}
         \caption{}
         \label{fig:ericssonpn}
     \end{subfigure}
   \begin{subfigure}[b]{0.44\textwidth}
         \centering
         \includegraphics[width=\textwidth]{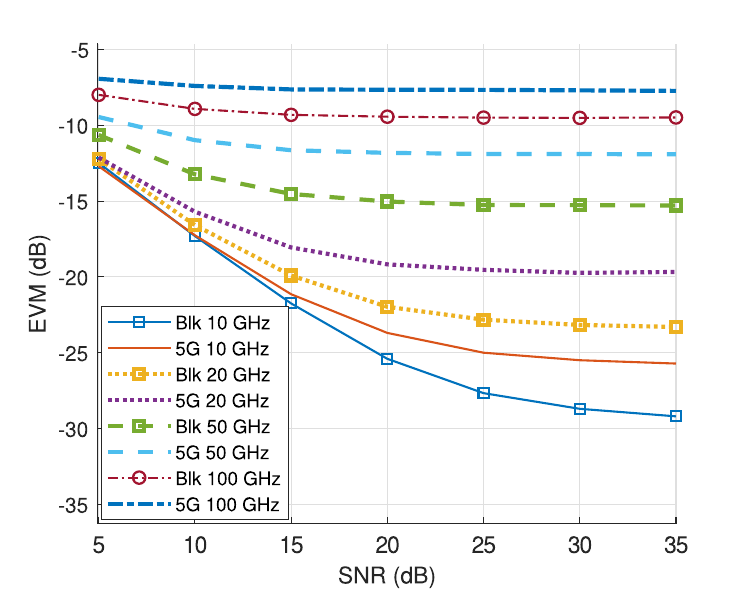}
         \caption{}
         \label{fig:ericssonevm}
     \end{subfigure}
    \caption{5G NR versus Block PTRS performance using phase noise model proposed by Ericsson \cite{R1-2003851}, at difference operation frequencies.}
    \label{fig:5GvsBlk}
\end{figure}

\begin{table}[h]
\caption{Simulation Parameters}
\label{tab:sim_params}
\centering
\begin{tabular}{|l|l|}
\hline
\textbf{Parameter} & \textbf{Value} \\
\hline
Subcarrier spacing & 15 kHz \\
Number of RBs & 10 \\
Total subcarriers & 120 \\
5G NR PTRS configuration & 8 groups, 4 PTRS per group \\
Block PTRS configuration & 88 data + 32 PTRS  \\
IFFT size & 1024 \\
Sampling rate & 1/(15×10³×1024) seconds \\
PTRS amplitude & $1/\sqrt{2}$ of max. data\\
Phase noise model & Ericsson model \\
Center frequencies & 10, 20, 50, 100 GHz \\
SNR range & 5 to 35 dB \\
Channel type & AWGN (single tap) \\
Receiver PN interpolation & Spline \\
5G NR DFT size & 120 \\
Block DFT$_1$ size & 88 \\
Block DFT$_2$ size & 32 \\
CP length & Normal \\
Channel estimation & Perfect \\
\hline
\end{tabular}
\end{table}

The simulation framework implements both the conventional 5G NR PTRS and our proposed block-based approach. The PTRS amplitude is scaled to $1/\sqrt{2}$ of the maximum constellation amplitude, ensuring consistent power allocation across different modulation schemes while maintaining robust phase-tracking performance. We assume a single-tap AWGN channel with perfect channel estimation and apply spline-based receiver interpolation to mitigate phase noise effects.

In Fig. \ref{fig:trackingsnapshot}, we present a snapshot of the phase-noise tracking performance for both the 5G NR and proposed block-based schemes. In each scheme, the phase noise is first estimated at the group centers and averaged over the group samples. Spline interpolation is then applied to estimate phase noise at intermediate points. We observe that, at 10 GHz, the block-based scheme—which employs 32 estimation points—more closely follows the noise variations than the 8-point 5G NR configuration. In addition, the block-based approach filters each estimate over a bandwidth of 32 frequency bins, whereas the 5G NR scheme spans 120 bins and thus requires a grouping strategy to manage the faster fluctuations. The corresponding EVM results for both schemes are presented next.

\begin{figure}[h]
     \centering
     \includegraphics[width=.45\textwidth]{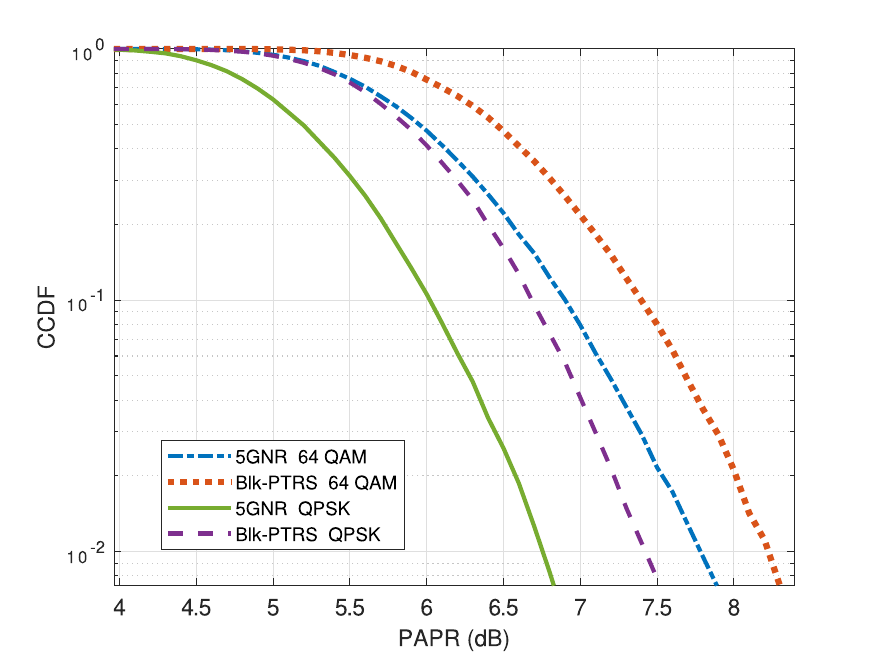}
     \caption{PAPR evaluation of Block vs. 5G NR PTRS allocation for QPSK and 64 QAM modulations.}
     \label{fig:trackingsnapshot}
 \end{figure}
The EVM of the data symbols after phase-noise correction is computed using
\begin{equation}
    EVM_{dB} = 10~ log_{10} \left(  \frac{\sum_{n=1}^{N_{data}} |{s_n - \hat s_n }|^2 }{\sum_{n=1}^{N_{data}} |{s_n }|^2} \right),
\end{equation}
where $s_n$ is the nth transmitted data symbol, $\hat s_n$ is the received corrected symbol. and the EVM Normalization Reference is set to Reference RMS.  
The results are presented in Figure 14. For reference, the Ericsson phase-noise spectrum model is plotted in Fig. 14a, illustrating how the phase noise increases with center frequency. In turn, this trend is evident in the EVM plots of Fig. 14b, where there is roughly a 3 dB gain at 100 GHz, and the improvement is even more pronounced at lower frequencies such as 10 GHz.

Before concluding, we also address the peak-to-average power ratio (PAPR). Because the proposed block-based method adds a parallel SC-OFDM PTRS stream to the existing data SC-OFDM symbols, the overall signal PAPR may increase. Fig. 15 shows a modest rise in PAPR by about 0.6 dB for lower-order modulation, which decreases to 0.4 dB for higher modulation orders.


\section{Conclusion}


This paper has introduced an innovative block PTRS allocation scheme for DFT-s-OFDM systems designed to enhance phase noise compensation by leveraging modulated reference symbols in the pre-DFT domain. Unlike the traditional approach outlined in the 5G-NR standard, which employs PTRS subcarriers within the data symbols for phase noise estimation, our method positions a distinct block of PTRS in parallel with the data symbols. 
Additionally, we have delineated the modulation and demodulation framework alongside a receiver block dedicated to phase noise estimation and correction. 

Through analysis and simulations, we demonstrated the superior performance of our proposed block PTRS allocation compared to the conventional 5G-NR method. Specifically, our results show a significant reduction in rms phase error and EVM, particularly in scenarios characterized by lower noise floors. For example, in our simulations with phase noise amplitude $a=0.5$ and noise floor $nf=2\times10^{-3}$, the block PTRS allocation achieved an rms phase error reduction of 18\% compared to the 5G-NR allocation.

Moreover, our study highlighted the importance of the trade-off between the repetition rate and the pulse bandwidth in mitigating phase noise effects. The use of sinc-shaped pulses in the block PTRS configuration proved to be more effective in reducing phase noise, as evidenced by higher performance gains across various phase noise spectrums. In contrast, the raised cosine-shaped pulses (roll-off factor $\alpha=1$), despite their wider bandwidth, did not perform as well due to using a lower repetition rate.

Finally, the performance of block PTRS was compared with 5G NR PTRS allocation, showing superior performance in the presence of phase noise model at 10, 20, 50, and 100  GHz, at the expense of 0.6dB higher PAPR.   
Finally, the performance of block PTRS was compared with the 5G NR PTRS allocation, showing superior performance in the presence of the Ericsson phase noise model at 10, 20, 50, and 100 GHz, albeit with a 0.6 dB increase in PAPR at a CCDF of $10^-2$. These findings underscore the effectiveness of block PTRS in challenging phase noise environments, despite the slight PAPR penalty.

\section*{Acknowledgment}
This material is based upon work supported in part by the U.S. Department of Energy, Office of Science, Office of Advanced Scientific Computing Research, Early Career Research Program under Award Number DE-SC-0023957, and in part by the National Science Foundation under Grant No. 2323300.
\ifCLASSOPTIONcaptionsoff
  \newpage
\fi


\bibliographystyle{IEEEtran}
\bibliography{IEEEabrv,PN}

\end{document}